\documentclass[proof]{WileyASNA-v1}

\articletype{Article Type}%

\usepackage [latin1]{inputenc}
\received{02 april 2021}
\revised{11 april 2021}
\accepted{15april 2021}

\begin{document}

\title{Pushing the limits of  time beyond the Big Bang singularity: The branch cut universe}

\author[1,2]{C\'esar A. Zen Vasconcellos*}

\author[3,4]{Peter O. Hess}

\author[1]{Dimiter Hadjimichef}

\author[5]{Benno Bodmann}

\author[6]{Mois\'es Razeira}

\author[1]{Guilherme L. Volkmer} 

\authormark{C\'esar A. Zen Vasconcellos, Peter O. Hess, Dimiter Hadjimichef, Benno Bodmann, Mois\'es Razeira, and Guilherme L. Volkmer }

\address[1]{\orgdiv{Instituto de F\'isica}, \orgname{Universidade Federal do Rio Grande do Sul (UFRGS)}, \orgaddress{\state{Porto Alegre}, \country{Brazil}}}

\address[2]{\orgdiv{International Center for Relativistic Astrophysics Network (ICRANet), Pescara, Italy}}  

\address[3]{Universidad Nacional Aut\'onoma de Mexico (UNAM), M\'exico City, M\'exico}

\address[4]{Frankfurt Institute for Advanced Studies (FIAS), J.W. von Goethe University (JWGU), Hessen, Germany}

\address[5]{\orgdiv{Unversidade Federal de Santa Maria (UFSM), Santa Maria, Brazil}}

\address[6]{\orgdiv{Laborat\'orio de Geoci\^encias Espaciais e Astrof\'isica (LaGEA)}, \orgname{Universidade Federal do PAMPA (UNIPAMPA)}, \orgaddress{\state{Ca\c{c}apava do Sul}, \country{Brazil}}}

\corres{*Av. Bento Gonçalves, 9500 - Agronomia, Porto Alegre - RS, 91501-970.  \email{cesarzen@cesarzen.com}}

\abstract{In this article we follow a previously developed theoretical approach, based in the tools of the singular semi-Riemannian geometry, to push the limits of time beyond the primordial spacetime singularity.  By complexifying the Friedmann-Lema\^itre-Robertson-Walker (FLRW) metric
and Friedmann's equations we model a branch cut universe,  in which the  cosmic FLRW metric scale factor is analytically continued to the complex plane, and becomes equivalent from a conceptual point of view of describing a hypothetical general metric of maximally symmetric and homogeneous superposed multiple universes. 
} 

\maketitle

\section{Introduction}

In the standard cosmological model of general relativity~\citep{Einstein1916}, described by the
Friedmann-Lema\^itre-Robertson-Walker (FLRW) metric~\citep{Friedmann1922}-\citep{Walker1937}, 
Friedmann's equations represent a closed set of equations which relate the scale factor $a(t)$, the energy density $\rho(t)$ and the pressure $p(t)$ for a flat,
open and closed universe.

In a search to overcome the presence of singularities in general relativity, we have combined in a recent publication~\citep{Zen2020} the multiverse proposal by S. Hawking and T. Hertog of a hypothetical set of multiple universes, existing in parallel~\citep{Hawking2018} and the technique of analytical continuation 
 in complex analysis applied to the Friedmann-Lema\^itre-Robertson-Walker (FLRW) metric. As a result we obtained Friedmann-type equations for the a complex version of the $\Lambda$CDM  ($\Lambda\neq 0$) model:
\begin{eqnarray}
H_{\xi}(t)&=&\frac{8\pi\,G}{3\,}\rho_{\xi}-\frac{k\,}{a_{\xi}^2}+\frac{\Lambda_{\xi}\,}{3} \label{H}
\\
2\frac{\ddot{a}_{\xi}}{a_{\xi}}&=&-{8\pi\,G}\,p_{\xi}-H_{\xi}(t) -\frac{k\,}{a_{\xi}^2}+\Lambda_{\xi}\,, \label{a}
\end{eqnarray}
where $H_{\xi}(t)=\dot{a}^{2}_{\xi}(t)/a^{2}_{\xi}(t)$, and with the scale factor $a_{\xi}(t)$ assumed to be analytically continued to the complex plane~\citep{Zen2020}. 

The deduction of the complex Friedmann's-type equations was carried out through the following simple canonical steps~\citep{Zen2020}: 
\begin{enumerate}
\item[(a)] Assuming the multiverse conception and the superposition principle for linear independent systems,  we summed the resulting set of Friedman's-type equations for the complex $\Lambda$CDM  model on the parameter $\xi$ which scans the hypothetical set of multiple universes:
\begin{equation}
\sum_{\xi} H_{\xi}(t)= \sum_{\xi} \Biggl( \frac{8\pi\,G}{3\,}\rho_{\xi}-\frac{k\,}{a_{\xi}^2}+\frac{\Lambda_{\xi}\,}{3} \Biggr)\, ;  \label{H}
\end{equation}
\begin{equation}
2 \sum_{\xi} \frac{\ddot{a}_{\xi}}{a_{\xi}}=- \sum_{\xi} \Biggl( {8\pi\,G}\,p_{\xi}-H_{\xi}(t) -\frac{k\,}{a_{\xi}^2}+\Lambda_{\xi} \Biggr) \,, \label{a}
\end{equation}
where $H_{\xi}(t)=\dot{a}^{2}_{\xi}(t)/a^{2}_{\xi}(t)$. This formulation, with a complex cosmic scale factor $a_{\xi}(t)$ becomes equivalent from a conceptual point of view of describing a hypothetical general metric of maximally symmetric and homogeneous superposed multiple universes. Following this methodology, we have obtained a closed set of field equations with multiple singularities that relate the scale factor $a_{\xi}(t)$, the energy density $\rho_{\xi}(t)$ and the pressure $p_{\xi}(t)$ for a flat, open and closed universe, 
which reduce, similarly to the case of a single-pole metric, to 
  \begin{eqnarray}
\sum_{\xi}   \Biggl[   3 \Biggl( \frac{ \rho_{\xi}(t) + p_{\xi}(t)}{a_{\xi} (t)} \Biggr)  + \frac{\dot{\rho}_{\xi}(t)}{\dot{a}_{\xi} (t)}  \Biggr]  & = & 0
\, . \label{NewFE}
\end{eqnarray}
Caution should be taken here. These equations are {\it not} a simply direct generalisation of the conventional Friedmann's equations based on the real FLRW single-pole metric. Nor a simple parameterization of $a(t)$. Due to the non-linearity of Einstein's equations, such a direct generalisation would not be formally consistent. The present formulation 
as stated earlier, is the outcome of complexifying the FLRW metric and results in a sum of equations associated to infinitely many poles (in tune with Hawking's assumption of infinite number of primordial universes that occurred simultaneously) arranged along a line in the complex plane with infinitesimal residues (for the details see~\citet{Zen2020}). The multiverse conception corresponds here to a theoretical mathematical device for implementation of the proposal.
\item[(b)]
To push the limits of the Friedmann's field equations beyond the primordial singularity, we shifted the variable $a_{\xi}(t)$ to $a_{\xi}(t) - \chi_{\xi}(t)$, where $\chi_{\xi}(t)$ represents a regularisation variable\footnote{The introduction of a regularisation function at this stage of the formulation is not equivalent to changing the limits of the integration of Friedmann's equations to avoid the presence of singularities. This is because essential or real singularities at $t = 0$ cannot be removed simply by any coordinate transformation. The technical procedure adopted here results in solutions conformed by branch cuts that allow to circumvent the singularities, which in turn become branch points. This procedure allows a formal treatment consistent with the Planck scales that establish, according to the multiverse concept, the region of confluence between quantum mechanics and general relativity.} extending from the Planck time $t_P$ to the present time $t$; the regularisation function allows the contour solution-lines to move around the branch cut, since the integration limits can be shifted without altering the continuity of the resulting functions so long as the contour-lines does not cross the complex branch-point related to the branch-cut.
\item[(c)]
Imposing that the multiple singularities of the field equations are confined to the {\it same} universe and using a Riemann sum to approximate equation (\ref{NewFE}) to a definite integral\footnote{Which implies the disappearance of the scanning index  $\xi$ of the multiple universes on the continuous variables $\chi_{\xi}(t)$, 
$a_{\xi}(t)$,  $\rho_{\xi}(t)$, and $p_{\xi}(t)$.}, we integrated the resulting equation in terms of the continuous variable $\chi(t)$:
\begin{equation}
   \int_{- \chi(t)}^{\chi(t)}  \Biggl[  3 \Biggl( \frac{ \rho(t) + p(t)}{a(t) - \chi(t)} \Biggr)  + \frac{\dot{\rho}(t)}{\dot{a}(t) - \dot{\chi}(t) }  \Biggr]  d\chi  =  0
\, .
\end{equation}
This equation may be rewritten in the following form:
\begin{equation}
   \Biggl[  \Bigl(\rho(t) + p(t) \!  \Bigr)   + \dot{\rho}(t)  \, \Bigl(\frac{d}{dt}\Bigr)^{-1} \Biggr] \int_{-\chi(t)}^{\chi(t)}   \frac{d\chi}{a(t) - \chi(t)}  = 0 \, , \label{RHS}
 \end{equation}
 where $ \Bigl(\frac{d}{dt}\Bigr)^{-1}$ acts on the inverse expression of the RHS of equation (\ref{RHS}).
 Integrating the right side part of this equation, results
\begin{equation}
 \Biggl[3 \Bigl(  \rho(t)   +  p(t)  \Bigr)   +  \dot{\rho}(t) \Bigl(  \frac{d}{dt}\Bigr)^{-1}  \Biggr]  \, \ln \Biggl(\frac{a(t)  +  \chi(t)}{a(t)  -   \chi(t)} \Biggr) = 0  \, .  \label{del}
 \end{equation}
  We define:
\begin{equation}
\beta(t) \equiv \frac{a(t)  +  \chi(t)}{a(t) - \chi(t)}  \, .
\end{equation}
 To perform the derivative of the rhs of (\ref{del}), we use :
 \begin{eqnarray}
 \frac{d}{dt} \ln^n(\beta(t) & = & n  \, \ln^{n-1}[\beta(t)] \frac{d}{dt} \ln[\beta(t)] \nonumber \\
 & = & n  \, \ln^{n-1}[\beta(t)] \frac{1}{\beta(t)} \dot{\beta}(t).  \label{del2}
 \end{eqnarray}
 Combining (\ref{del}) and (\ref{del2}), the previous expression reduces to
\begin{equation}
 \Longrightarrow 3 \Bigl( \rho(t) + p(t)  \Bigr) 
   -   \dot{\rho}(t) \, \ln[\beta(t)] \frac{\beta(t)}{\dot{\beta}(t)}
  =  0 . 
   \label{FEi}
\end{equation}

This expression corresponds to a Friedmann's-type equation with a cut from $-\chi(t)$ to $\chi(t)$ for a variable value of $t$.  In the following, we seek for solutions of Friedmann's equations for a branch cut universe.

Equation (\ref{FEi}) may be rewritten for the radiation-dominated era as  (with $p=\rho/3$)
\begin{eqnarray} 
 &&  \ln[\beta(t)]\frac{{\beta}(t)}{\dot{\beta}(t)}  =   \frac{4 \rho(t)}{\dot{\rho}(t)} 
    \Rightarrow \ln^{-1}[\beta(t)] \frac{\dot{\beta}(t)}{\beta(t)}  =    \frac{\dot{\rho}(t)}{4 \rho(t)} \, ,   \nonumber    \\
&&   \Rightarrow -4 \ln^{-5}[\beta(t)] \frac{\dot{\beta}(t)}{\beta(t)}\rho(t)  + \ln^{-4}[\beta(t)]{\dot{\rho}(t)}  =    0 \, ,  \label{FEi2} \end{eqnarray}
so, the solution of equation (\ref{FEi2}) is
\begin{equation}
\frac{d}{dt} \biggl[  \ln^{-4}[\beta(t)] \, \rho(t)      \Biggr] = 0 \, .
\end{equation}
From this equation we have
\begin{equation}
 \ln^{-4} \bigl[ \beta(t) \bigr] \, \rho(t)   =  constant  \, ,
\end{equation}
whose solution can be written in a convenient way
\begin{equation}
 \ln^{-4} \bigl[ \beta(t) \bigr] \, \rho(t)   =  \ln^{-4} \bigl[ \beta(t_0) \bigr] \, \rho(t_0)     \, , 
\end{equation}
which leads to
\begin{equation}
 \fbox{$\rho(t)   =  \ln^{4} \bigl[ \beta(t) \bigr]  \ln^{-4} \bigl[ \beta(t_0) \bigr] \, \rho_0$}     \, . \label{1}
\end{equation}
The corresponding solutions for the matter-dominated era is (with $p = 0$):
\begin{equation}
\ln^{-3} \bigl[ \beta(t) \bigr] \, \rho(t)  =  \ln^{-3} \bigl[ \beta_0 \bigr] \, \rho_0 \, , 
\end{equation}
or equivalently as
\begin{equation}
\fbox{$\rho(t) =  \ln^{3} \bigl[\beta(t) \bigr] \ln^{-3} \bigl[ \beta(t_0) \bigr]  \, \rho_0$} \, . \label{2}
\end{equation}
For comparison, we present below the corresponding classical results of the scale factor solutions using the conventional FLRW metric, which shows the consistency of the proposal:
\begin{eqnarray}
a^{4}(t) \rho(t)  =  a^{4}(t_0) \rho_0
\Rightarrow   \fbox{$\rho(t) = a^{-4}(t) \, a^{4}(t_0) \rho_0 $} \, , \label{3} \\
a^{3}(t) \rho(t)  =  a^{3}(t_0) \rho_0 
 \Rightarrow   \fbox{$\rho(t) = a^{-3}(t) a^{3}(t_0) \rho_0$} \, . \label{4}
\end{eqnarray}
\end{enumerate}

\subsection{About the new scale factor}

 In the FLRW metric, $ a (t) $ represents a real and dimensionless scale factor which characterises the expansion of a homogeneous, isotropic, single-pole, expanding and path-connected universe. 

Here, $\ln^{-1}[\beta(t)]$ represents as stressed before a complex scale factor of  a hypothetical general metric of maximally symmetric and homogeneous superposed multi-pole expanding universes existing in parallel, following the multiverse conception that explore points of confluence between quantum mechanics and general relativity. As emphasised, our proposal assumes the `multi-universes' are infinitesimally separated and that the multiple singularities of the field equation are confined to the {\it same} universe, an assumption that originates a branch-cut universe.
 
 \subsubsection{\lowercase{$\ln[\beta(t)]$} as a scaling factor in time} 
 
  From equations (\ref{1}), (\ref{2}), (\ref{3}), and (\ref{4}) we conclude that the original scale of evolution of the universe, $a(t)$, is replaced here by a new scale $\ln^{-1}[\beta(t)]$ and that the density of the universe presents the following scaling functional form considering the two formulations:
\begin{equation}
\rho^{1/n}(t) \propto  
\Biggl\{
\begin{array}{c}
  a^{-1}(t)   \\ \\
   \ln[\beta(t)]  \\  
\end{array} \,\,; \, \, \, n = 3,4
\end{equation}
Thus, the density of the universe scales in the FLRW metric formulation as 
 $\rho^{1/n}(t) \propto 1/a(t)$, while in the analytically continued formulation, the density of the universe scales as $\rho^{1/n}(t) \propto \ln[\beta(t)]$, with $n=3,4$. 
 
The analysis of the new scale factor, $\ln^{-1}[\beta(t)]$ brings also a crucial aspect, as follows. 
The relation of the proper time $\tau$ of a co-moving system and the parametric time $t$ is given conventionally, 
in differential form, as  
\begin{equation}
d\tau = dt/a(t)  \, ,
\end{equation} 
which is mapped in the present formulation as
\begin{equation}
d\tau = \ln[\beta(t)] dt   .
\end{equation}
As a result, we conclude that the inverse of the new complex scale factor, i.e. $\ln[\beta(t)]$, represents  a linear {\it scaling factor in
time}, bringing to time a complex nature, with an imaginary component\footnote{As historically stated by~\citet{Minkowski1915}, in general relativity and by~\citet{Matsubara1955}, in statistical mechanics.}.

\begin{figure}[htbp]
\centering
\includegraphics[width=43mm,height=50mm]{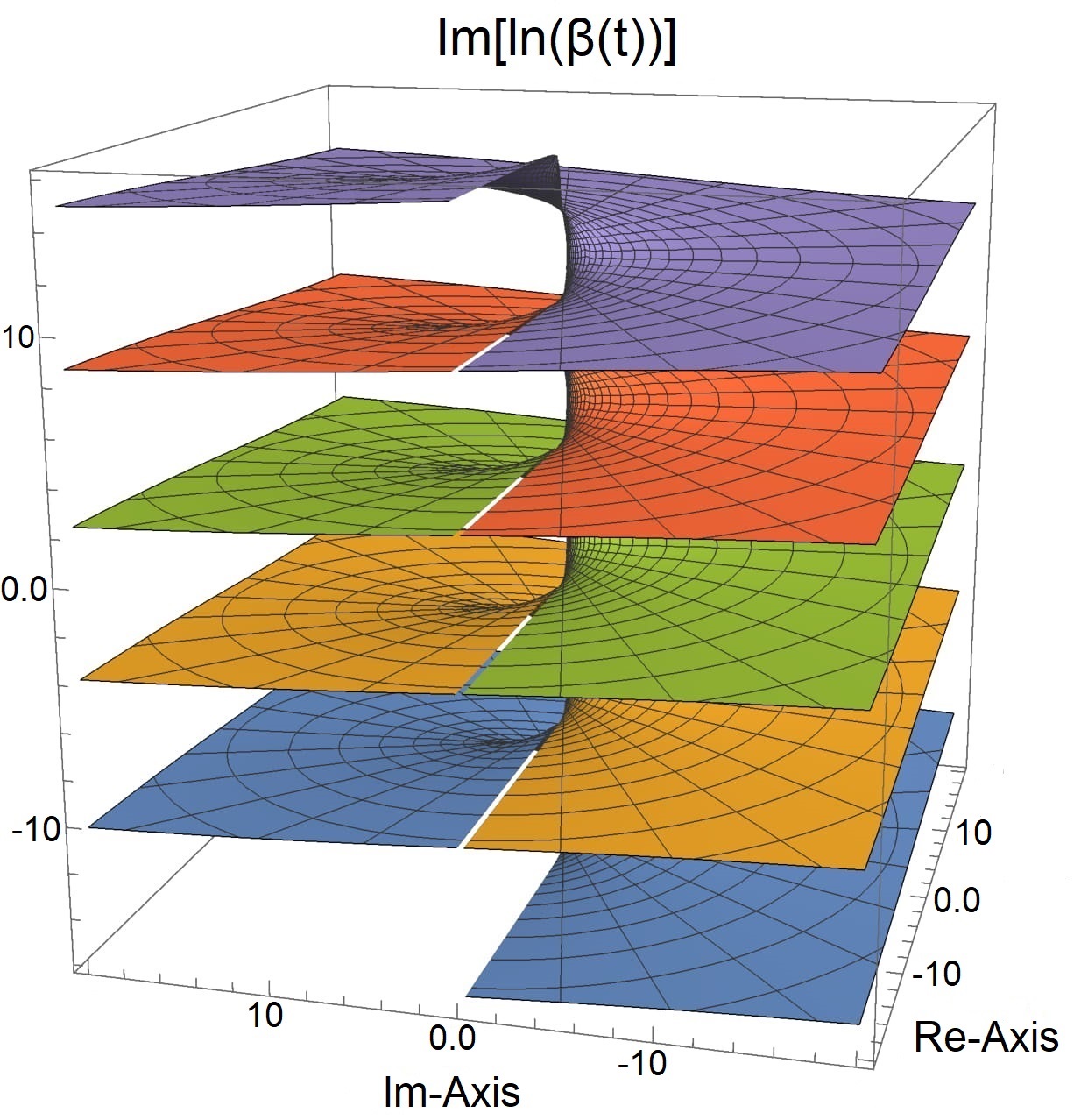} 
\includegraphics[width=43mm,height=50mm]{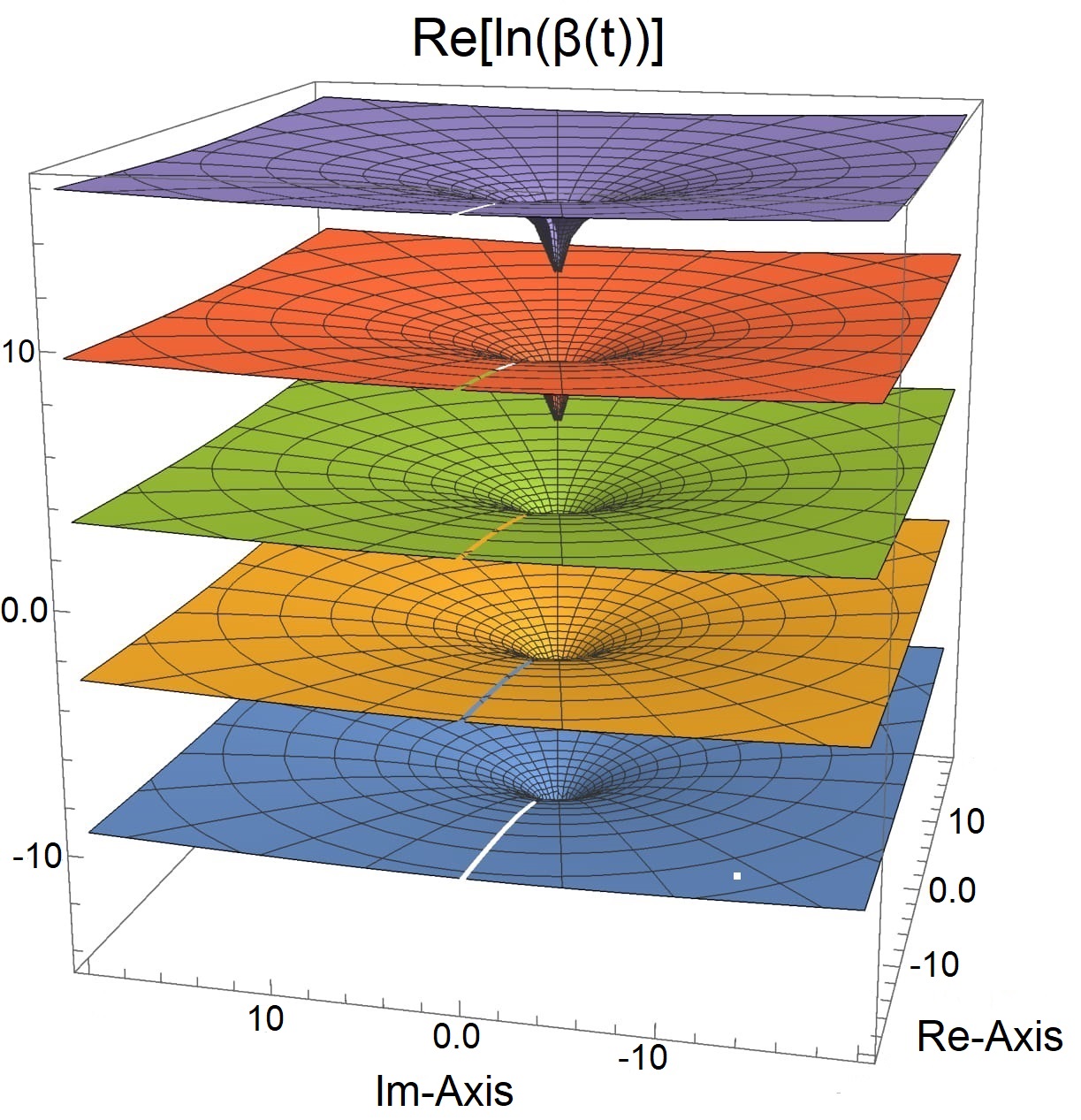} 
\caption{
Left figure: characteristic plot of the Riemann surface $R$ associated with the imaginary part of the scaling factor in time, $\ln[\beta(t)]$, represented by $Arg(\beta(t))$. The figures shows the resulting connected glued domains: the various branches of the function are {\it glued} along the copies of each upper half plane with their copies on the corresponding lower half planes.  Each two copies can be visualised as two levels of a {\it continuously spiralling parking garage}, from ``level" $\ln z = \ln \kappa(r) + i \theta $ for instance to the ``level" $\ln z = \ln \kappa(r) + i (\theta + 2 \pi) $ or to the ``level $ \ln z = \ln \kappa(r) + i (\theta - 2 \pi) $, and so on. As a final result we have a connected Riemann surface with infinitely many ``levels",  $ \ln z = \ln \kappa + i (\theta \pm 2 n \pi) $, extending clockwise or counterclockwise both upward and downward. For simplicity, the design is limited to a few Riemann sheets. 
The important transition region corresponds however to the domain that brings together the principles of quantum mechanics and gravity general relativity.  Right figure: characteristic plot of the real part of $\ln[\beta(t)]$, which corresponds to $\tau = |\beta (t)| = \sqrt{\kappa^2_{x} + \tau^2_{y}}$ decomposed in two components in the form $\kappa = (\kappa_{x},\kappa_{y})$ (in a temporal scale of billions of years) and shows a set of multiverses. This procedure allows obtaining complex solutions of Friedmann's-type integral equations  of an evolutive universe in which the spacetime fabric develops continuously along Riemann sheets that circumvent the branch cut, thus avoiding discontinuations of the general relativity equations. The corresponding solutions describe a {\it branch cut universe}, with a cut from $-\chi(t)$ to $\chi(t)$,  which can be thought as stressed before as a sum of infinity single-poles arranged along a line in the complex plane with infinitesimal residues.
} \label{fig1}
\end{figure}

In figure (\ref{fig1}), we present a characteristic plot of the scaling factor in time $\ln[\beta(t)]$  of the branch cut universe. 

\begin{figure}[htbp]
\centering
\includegraphics[width=43mm,height=50mm]{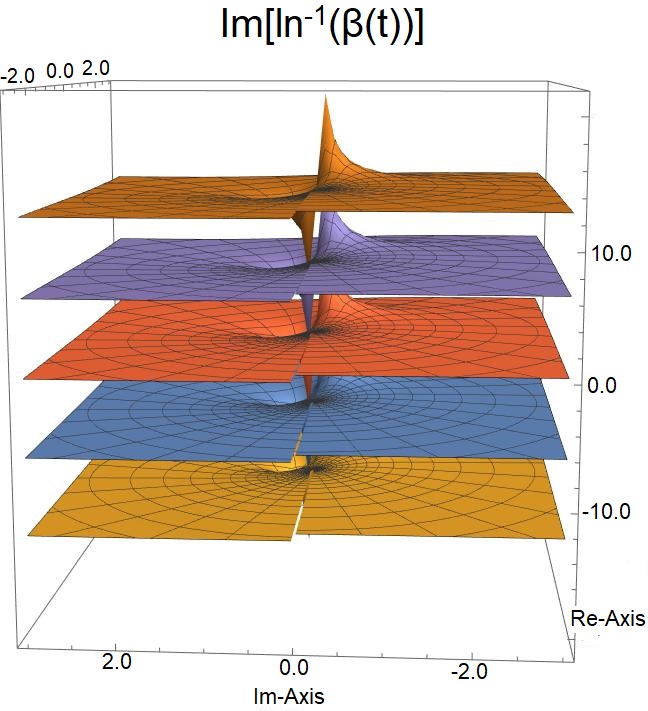}
\includegraphics[width=43mm,height=50mm]{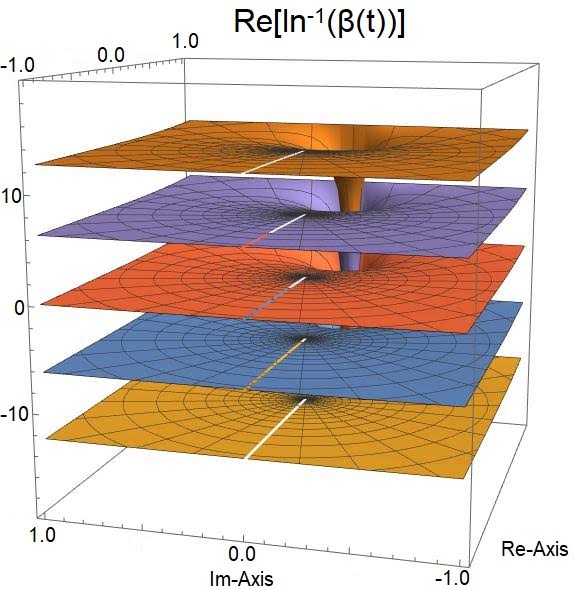}
\caption{
Left figure: characteristic plot of the Riemann surface $R$ associated to the imaginary part of the scale factor of the analytically continued FLRW metric $\ln^{-1}[\beta(t)]$. Right figure: characteristic plot of the real part of $\ln^{-1}[\beta(t)]$.} \label{fig2}
\end{figure}

\subsubsection{Geometric meaning of \lowercase{$\ln^{-1}[\beta(t)]$}}

In the following, we analyze the physical and geometric meaning of the new scale parameter.
The analytically continued
Ricci scalar, $R_{[\rm{ac}]}$
\begin{equation}
R_{[\rm{ac}]}  = g_{[\rm{ac}]}^{\mu \nu} R_{[\rm{ac}] \mu \nu} \, , \label{Ricciac}
\end{equation}
where $R_{[\rm{ac}]\mu \nu}$ defines the analytically continued ([\rm{ac}]) Ricci curvature tensor $T_{[\rm{ac}]}$ becomes
\begin{eqnarray}
R_{[\rm{ac}]}  & \! = \! &  g_{[\rm{ac}]}^{\mu \nu} R_{[\rm{ac}]\mu \nu}  \label{RicciScalarac}   \\
& \! = \! &  6 \Biggl[ \! \Biggl(\! \frac{\frac{d^2}{dt^2}{\ln^{-1}[\beta(t)]}}{\ln^{-1}[\beta(t)]} \! \Biggr) \! + \! \Biggl( \! \frac{\frac{d}{dt}{\ln^{-1}[\beta(t)]}}{\ln^{-1}[\beta(t)]} \!  \Biggr)^2
 \! + \! \frac{k}{\ln^{-2}(\beta(t))} \! \Biggr] \, . \nonumber
\end{eqnarray}
From this expression, we additionally conclude that the new scale factor, $ \ln^{-1}[\beta(t)]$,   the solely dynamical degree of freedom in the analytically continued FLRW metric, 
shapes the curvature of a hyphothetical universe with characteristic parameters analytically continued to the complex plane. And unlike general relativity, the analytically continued Ricci scalar curvature do not bend to infinity at the Planck scale, thus eliminating, on the complex plane, the presence of essential singularities. 

On a quantum cosmology formulation, the new scale factor, $\ln^{-1}[\beta(t)]$ represents a dynamical complex component to be quantised (for the details see~\citet{ZenB}). However, after the realisation of time through a Wick rotation, the complex feature disappears~\citep{ZenB}. 

\begin{figure}[h]
\centering
\includegraphics[width=45mm,height=70mm, angle =90]{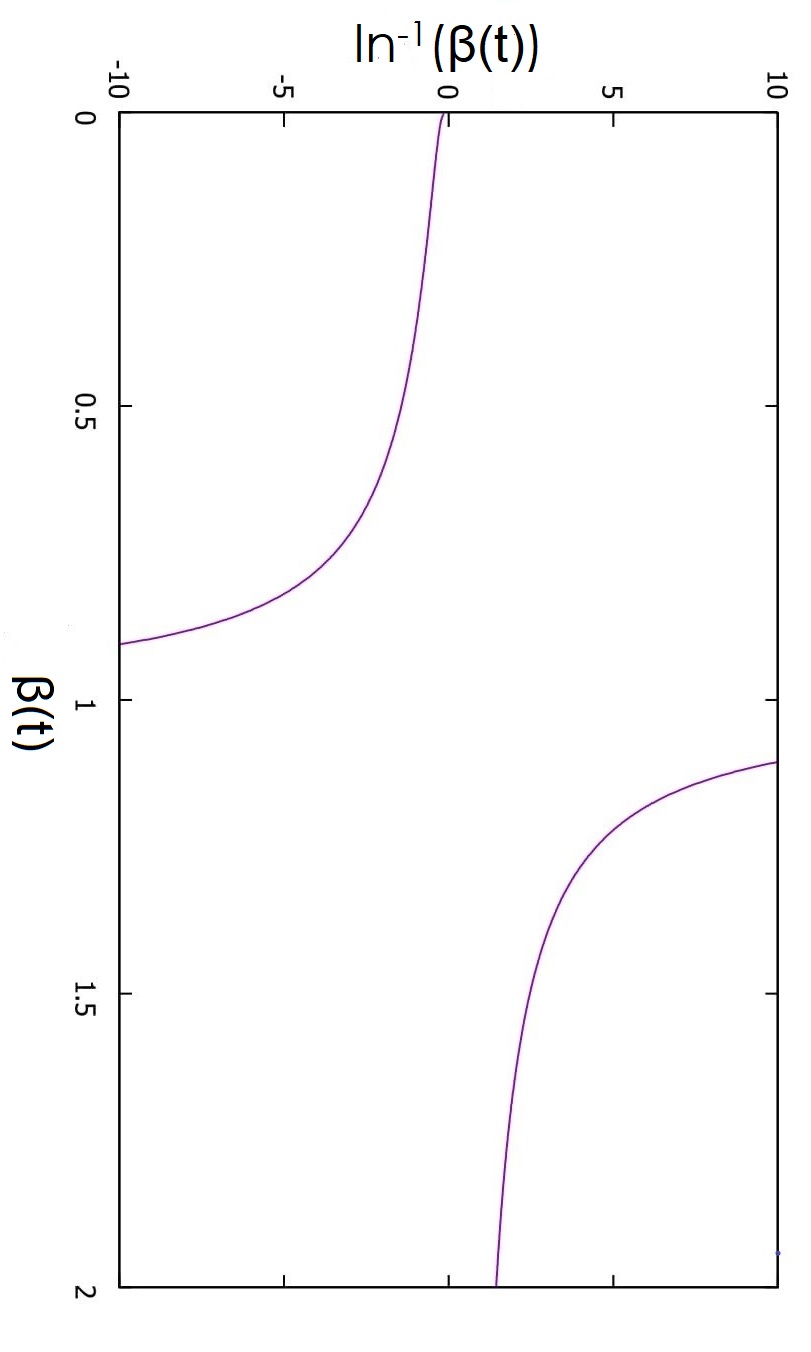} 
\caption{Plot of $\ln^{-1}[\beta(t)]$ as a function of $\beta(t)$. 
} \label{fig3}
\end{figure}

Additionally, the Ricci curvature scalar which characterises the radius of the universe, can be expressed, as a result of complexifying the FLRW metrics, in terms of the new scale factor $\ln^{-1}[\beta(t)]$); we represent this parameter in polar form as $\beta(t) =  \kappa(t) e^{i n \theta}$, with $\kappa(t)$ directly related to the complex analytically  continued Ricci scalar curvature and  to the complex analytically continued {\it radius} of the universe~\citep{Zen2020}. This representation allows to map the behaviour of the 
$\ln^{-1}[\beta(t)]$ parameter in terms of level curves that describe the slope and variations of a hypothetical topological contour, very useful in a mathematical analysis of the implications of the presence of a branch cut in the solutions of Friedmann's equations.  In a visualisation of $\ln[\beta(t)]$ shown in~\citet{Zen2020}, the Riemann surface appears to describe a spiral curve around a vertical line corresponding to the origin of the complex plane (see the left panel of Fig. (\ref{fig1})). In the right panel of Fig. (\ref{fig1}) 
we extended the graphical representation of the previous article showing the real part of this function, where some of the multiverses are represented. Evidently, the presence of multiverses in the real part is associated with the peculiar characteristics of the proposal in which the superposed multiple
universes as stressed before correspond to a convenient mathematical device. The representation of the real part of the new scale factor does not contain 
Riemann sheets,
the analytical varieties of complex dimension and therefore the `multi-universes' representation is still present, although with their singularities infinitesimally separated. In complex analysis, contour integrals are made in the complex sector, simply by summing the values of the complex residues inside the contour. 
In Fig. (\ref{fig2}) we plot the imaginary and real parts of $\ln^{-1}[\beta(t)]$ for which similar conclusions to the previous case can be drawn.
The actual surfaces of 
the images shown in Figs. (\ref{fig1}) and (\ref{fig2})
extends arbitrarily far both horizontally and vertically the representation sector. In Fig. (\ref{fig3}), the behaviour of $\ln^{-1}[\beta(t)]$ as a function of $\beta(t)$ is presented. 

\subsection{About the complex FLRW metric}

 In what follows we consider a standard transformation, `conformally equivalent' to the FLRW metric analytically continued to the complex plane, in accordance with  the condition\footnote{This transformation was adapted from the corresponding conformal transformation introduced by ~\citet{Clifton2012}.}
  \begin{equation}
 g_{[\rm{ac}]\mu \nu}(x) \equiv e^{\Gamma(x)} g_{\mu \nu}(x) \, , \label{ct}
 \end{equation}
 where $\Gamma(x)$ represents an arbitrary function of the space-time coordinates $x$. The corresponding line-element  $ds^2_{[\rm{ac}]\mu \nu}(x)$ and 
factor  $\sqrt{-g}_{[\rm{ac}]}$ are also transformed accordingly.

We assume the Lagrangian density of the scalar-tensor theory after the conformal transformation reads
\begin{eqnarray}
 {\cal L}    =  \frac{1}{16}  \sqrt{-g}  \Bigl(&& \!\!\!\!\! \!\!\! g(\phi) { \nabla}_{\mu} \phi^{\dagger}(x) {\nabla}^{\mu} \phi(x)  \label{Lphi} \\ & - &  h(\phi) m^2_{\phi} \phi^{\dagger}(x) \phi(x)  
 -  \Lambda(\phi)  -  f(\phi) R  \Bigr) , \nonumber
 \end{eqnarray}
where $f(\phi)$, $g(\phi)$,   $h(\phi)$ and $\Lambda(\phi)$ are arbitrary functions of the scalar-complex field
\begin{equation}
\phi(x) = \frac{1}{\sqrt{2}}\Bigl( \phi_1(x) + i \phi_2(x) \Bigr) \, , 
\end{equation}
 with a mass represented by $m_{\phi}$ and the real and imaginary components represented by independent real scalar fields $\phi_1$ and $\phi_2$. 

In case $f(\phi) = \Lambda(\phi) = 0$ and $g(\phi) = h(\phi) = 1$, the Lagrange density (\ref{Lphi}) becomes invariant under the global continuous $U(1)$ symmetry transformation
\begin{equation}
\phi(x) \rightarrow e^{i \alpha} \phi(x) \, ,
\end{equation}
where $\alpha$ is a constant in $\mathbb{R}$ and (in general) $e^{i \alpha} \in$ U(1). In accordance with Noether's theorem we may identify a conserved current $j^{\mu}$ and a Noether charge $Q$ associated to the Lagrange density (\ref{Lphi}):
\begin{equation}
j^{\mu} = -i \Bigl(\phi^{\dagger}(x) \partial^{\mu} \phi(x)  -  \bigl[\partial^{\mu} \phi^{\dagger}(x)\bigr] \phi(x)  \Bigr) \, , 
\end{equation}
\begin{equation}
Q =  \int d^3x j^0    \, .
\end{equation}

In the most general case, however, such underlying (continuous) symmetries and conservation laws are not necessarily obeyed.   In the following we assume
$f(\phi)$ and $\Lambda(\phi)$ are real functions and do not contain derivatives in $\phi(x)$.
The fields $\phi(x)$ and $\phi^{\dagger}(x)$ describe independent degrees of freedom with respective conjugated momenta given by:
\begin{equation}
\Pi(x) = \frac{\partial {\cal L}}{\partial (\partial_0 \phi(x))} = g(\phi) \dot{\phi}(x) \, ,
\end{equation}
and
\begin{equation}
\Pi^{\dagger}(x) = \frac{\partial {\cal L}}{\partial (\partial_0 \phi(x))} = g(\phi) \dot{\phi}^{\dagger}(x) \, .
\end{equation}
The corresponding Hamiltonian 
\begin{equation}
H  =  \int d^3 x \Bigl(g(\phi(x)) \bigl[ \Pi^{\dagger}(x)  \dot{\phi}^{\dagger}(x) + \Pi(x)  \dot{\phi}(x) \bigr]  
 - {\cal L} 
\Bigr) \, , 
\end{equation}
may be expressed as
\begin{eqnarray}
H  & = &  \int d^3 x   \Biggl( g(\phi)  \Bigl[ \dot{\phi}^{\dagger 2}(x)  +  \phi^{2}(x)+   {\bf \nabla} \phi^{\dagger}(t) \cdot {\bf \nabla}\phi(x)  \Bigr]  \nonumber \\
 & + & h(\phi) m^2_{\phi} \phi^{\dagger}(x) \phi(t) +  \Lambda(\phi) + f(\phi) R \Biggr) \, . 
\end{eqnarray}
Assuming coefficients $g$, $h$, and $f$ normalised to one, we may notice that:
\begin{enumerate}
\item[(a)] For $f(\phi)$ and $\Lambda(\phi) \geq 0$, since $R$ is positive definite:
\begin{enumerate}
\item[(i)] For $g = h = + 1$, the Hamiltonian is positive semi-definite and therefore bounded from
below;
\item[(ii)] For $g = h = - 1$, the Hamiltonian is negative semi-definite and therefore bounded
from above and $\phi(x)$ is a ghost field; 
\item[(iii)] For $g = - h$, 
the Hamiltonian is indefinite and so it is not bounded either from below
or from above; if $g = +1$ and $h = -1$, $\phi(x)$ represents a tachyon field. If $g = -1$ and $h = +1$, $\phi(x)$ represents a
tachyon ghost field. 
\end{enumerate}
\item[(b)] Other combinations of the parameters can induce (or not) the presence of ghosts and / or tachyons.
\end{enumerate}
For comparison see for instance~\citep{Conroy2017}.

\subsubsection{Ghost and tachyon criteria}

The graviton propagator, powered by a four-current $J_{\mu}(x)$, details the field propagation through space. Changes in the gravitational action may imply structural modifications of the propagator and the admission by the theory of states of negative energy (ghosts), generating instabilities, even at the classical level (Ostrogradksy instabilities); these perturbative instabilities may carry positive and negative energy modes~\citep{Conroy2017}.

In order to avoid the spectra of ghosts or tachyons, we may require the following conditions for a quantum field formulation of our analytically continued formulation:
\begin{enumerate}
\item Ghosts in relativity are physical excitations which come with a negative residue in the graviton propagator, so such a pole should not contain negative residues or ghosts. 
\item To avoid the presence of tachyons, the propagator of the $\phi(x)$ field must contain only first order poles at $k^2+m_{\phi}^2$ with real masses $m_{\phi}^2 \geq 0$, 

\end{enumerate}

General relativity is a ghost-free theory, that preserves unitarity. Ghosts arising in modified theories of general relativity are distinct from those emerging in the quantisation of non-abelian gauge theories (Faddeev-Popov ghosts)~\citep{Faddeev1967}. The latter are introduced in quantum field theories as `ingredients' of a path integral formalism to absorb unphysical degrees of freedom and are associated this way only with internal Feynman diagram lines. In the former, in turn, 
the appearance of ghosts is inevitable when higher-order derivative terms are introduced into the theory, except in the context of a perturbative approximation~\citep{Simon1991,Donoghue1994}.
 
\section{Analytically continued Hubble rate}

The new scale factor allows to define the  analytically continued Hubble rate $H_{\rm{ac}}(t)$ as
\begin{equation}
H_{\rm{ac}}(t)  \equiv \Biggl[  \frac{\frac{d}{dt} \ln^{-1}[\beta(t)]}{\ln^{-1}[\beta(t)]} \Biggr] \, .\label{Hr}
\end{equation}
From this expression, the time derivative of $H_{\rm{ac}}(t)$ gives
\begin{eqnarray}
\dot{H}_{\rm{ac}}(t) & =  & - H^2_{\rm{ac}}(t) \Biggl( 1  -  \frac{1}{H^2_{\rm{ac}}(t)} \Biggl[ \frac{\frac{d^2}{dt^2} \ln^{-1}[\beta(t)]}{\ln^{-1}[\beta(t)]}\!\Biggr]  \Biggr) \, , 
\nonumber \\ & \equiv & \! H^2_{\rm{ac}} (1 + q_{\rm{ac}}),
\end{eqnarray}
which leads to
\begin{equation}
 q_{\rm{ac}}  \equiv  - \frac{1}{H^2_{\rm{ac}}(t)} \Biggl[ \frac{\frac{d^2}{dt^2}\ln^{-1}[\beta(t)]}{\ln^{-1}[\beta(t)]} \Biggr] \, ;
\end{equation}
$q_{\rm{ac}}$ defines the analytically continued deceleration parameter which provides a relationship between the density of the
universe and the critical density ($\rho_{\rm{cr}}$), 
for radiation-dominated (\rm{RD}) and matter-dominated eras (\rm{MD}), in the form
\begin{eqnarray}
q_{\rm{ac}}^{\rm{RD}} & = & \frac{\rho(t)^{\rm{RD}}}{\rho_{\rm{cr}}^{\rm{RD}}} \Rightarrow \rho(t)^{\rm{RD}} = \frac{3H^2_{\rm{ac}}(t)}{8\pi G}q_{\rm{ac}}^{\rm{RD}}\, , \\
\mbox{and} \,\,\, q_{\rm{ac}}^{\rm{MD}} & = & \frac{\rho(t)}{2 \rho_{\rm{\rm{cr}}}^{\rm{MD} }} \Rightarrow \rho(t)^{\rm{MD}} = \frac{3H^2_{\rm{ac}}(t)}{4\pi G}q_{\rm{ac}}^{\rm{MD}} \, \, .
\end{eqnarray}

\section{Analytically continued Friedmann's equations}

In what follows, is important to distinguish between critical time ($ t_ {\rm{cr}} $), Planck time ($ t_ {P} $) and the time associated with the origin of the universe ($ t = 0 $) in the Big Bang model.
Moreover, $\rho_0 = \rho(t_{\rm{cr}})$ denotes the critical density of the universe, i.e. $
\rho_0 = \frac{3 H^2(t_{\rm{cr}})}{8 \pi G} \sim 10^{-29} g/cm^3  \, , $
and $\beta_0 = \beta(t_0) = \beta(t_{\rm{cr}})$ with
$t_{\rm{cr}}$ defining the {\it critical time}, i.e., the time for the matter density of the universe to become spatially flat. 

\begin{figure*}[htbp]
\centering
 \includegraphics[width=46mm,height=46mm, angle =00]{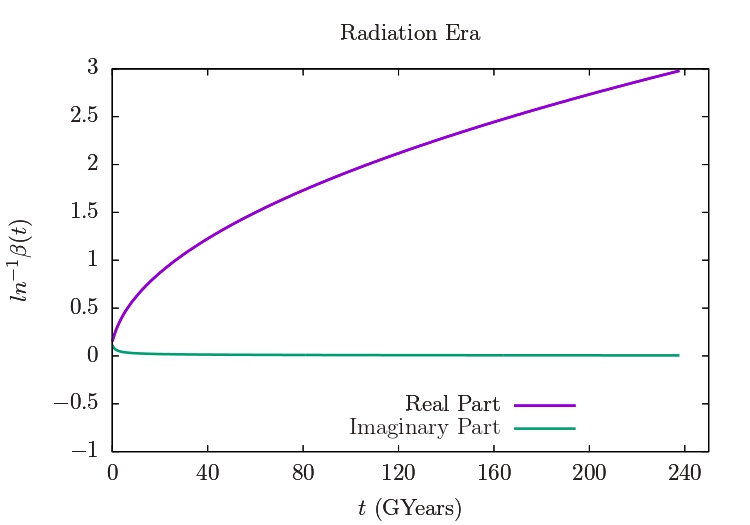} \hspace{-0.35cm}
\includegraphics[width=46mm,height=46mm, angle =00]{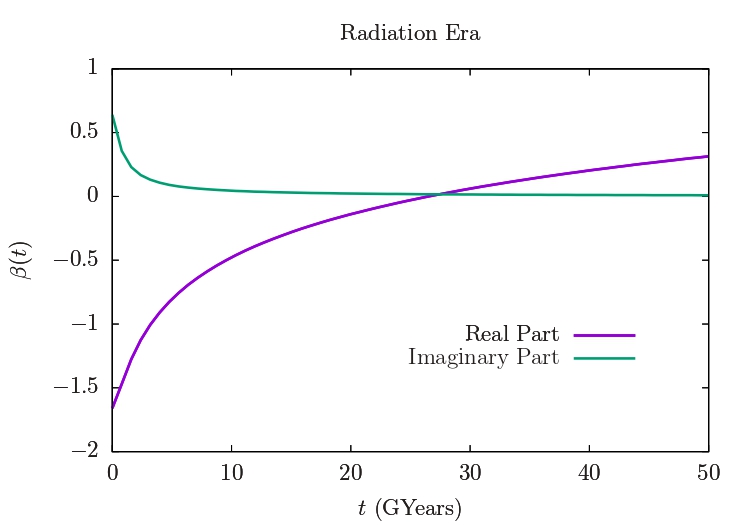} \hspace{-0.35cm}
\includegraphics[width=46mm,height=46mm, angle =00]{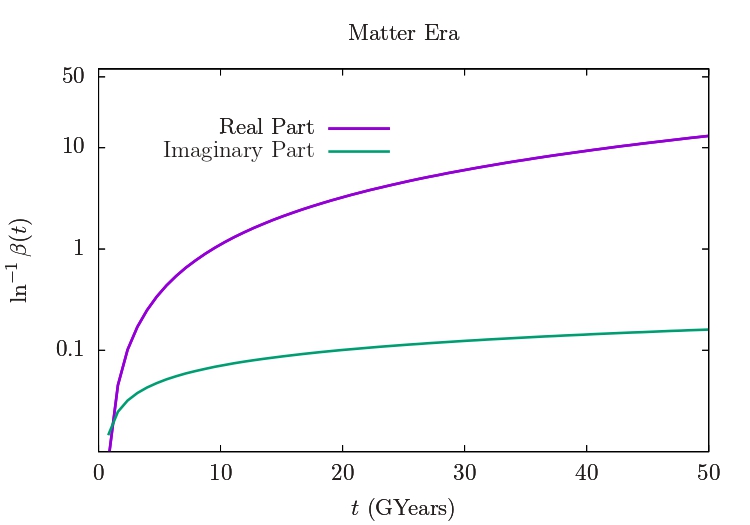} \hspace{-0.35cm}
\includegraphics[width=46mm,height=46mm, angle =00]{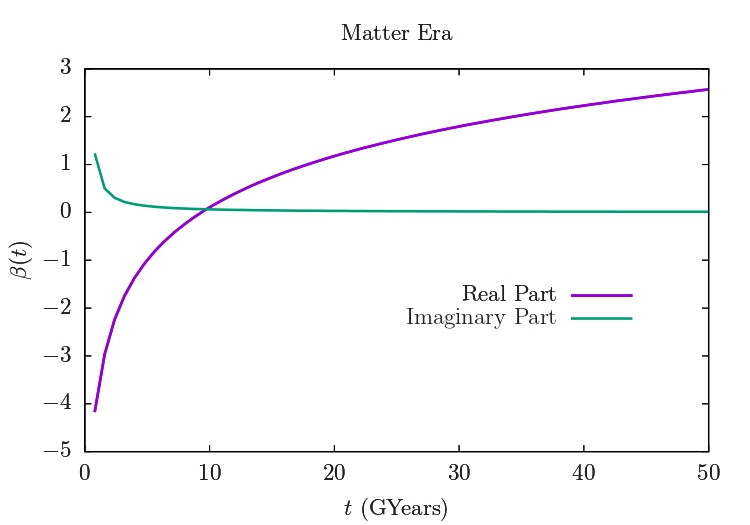}\caption{On the left: plot of the characteristic solutions of (\ref{eq1})  and (\ref{beta}) for the radiation-dominated era. On the right: plot of the characteristic solutions of (\ref{eq2}) and (\ref{beta2}) for the matter-dominated era.} \label{solacR}
\end{figure*}

From the previous equations, we obtain a new set of equations, for $k$ and $\Lambda$ different from zero and $c \neq 1$, 
\begin{equation}
\Biggl(\frac{\frac{d}{dt} \ln^{-1}[\beta(t)]}{\ln^{-1}[\beta(t)]  } \Biggr)^2   =    \frac{8 \pi G}{3} \rho(t) 
-  \frac{kc^2}{\ln^{-1}[\beta(t)]} + \frac{1}{3} \Lambda \, ,   \label{NFE1} 
\end{equation}
\begin{equation}
\Biggl( \frac{\frac{d^2}{dt^2} \ln^{-1}[\beta(t)]}{\ln^{-1}[\beta(t)] } \Biggr)   =  - \frac{4 \pi G}{3} \Big(\rho(t) + \frac{3}{c^2} p(t) \Bigr)
+  \frac{1}{3} \Lambda 
 ,  \label{NFE2}
\end{equation}
referred  as the first (\ref{NFE1}) and second (\ref{NFE2}) new Friedmann's-type  equations analytically continued to the complex plane (for comparison with the conventional treatment see~\citet{Bazin1965}),  and an
analytically continued energy-stress conservation law in the expanding universe
\begin{equation}
\frac{d }{dt}\rho(t) + 3 \Big(\rho(t) + \frac{p(t)}{c^2}  \Bigr) \Biggl( \frac{\frac{d}{d t} \ln^{-1}[\beta(t)]}{\ln^{-1}[\beta(t)]}\Biggr)  =  0 \, .  \label{ECln}
  \end{equation}%
In  appendix A we present mathematical solutions of the analytically continued Friedmann's equations for the different ages of the universe. In Fig. (\ref{solacR}) we plot the characteristic solutions of these equations for the radiation- and matter-dominated eras.

\subsection{Thermodynamics}%
 
The analytically  continued energy-stress conservation law (\ref{ECln}) may be written in the convenient form
 \begin{eqnarray}
&  \Bigl(\! \ln^{-1}[\beta(t)] \! \Bigr)^{\!3} \!\! d \! \rho(t)  \! + \! 3 \Big( \! \rho(t) \! + \! p(t)/c^2 \! \Bigr) \Bigl( \! \ln^{-1}[\beta(t)] \! \Bigr)^{\!2}  \! d  \! \ln^{-1}[\beta(t)]  &  \nonumber \\
&  \rightarrow d  \Bigl[ \! \rho(t) \Bigl(\! \ln^{-1}[\beta(t)] \! \Bigr)^{\!3} \! \Bigr]  \! + \!  (p(t)/c^2) \,  d  \Bigl( \! \ln^{-1}[\beta(t)] \!  \Bigr)^{\!3} = 0 \, . &\label{thermo}
 \end{eqnarray}
For any co-moving volume, the first term of the left expression of equation (\ref{thermo}) may be identified with
\begin{equation} 
\Bigg| 
 d \Bigl[\rho(t) \Bigl( \ln^{-1}[\beta(t)] \Bigr)^3 \Bigr] \Bigg| \propto  d E_M/c^2 = d M \, , 
\end{equation}
where  $d M$ represents an elementary relativistic analytically  continued mass-energy quantity
contained in the volu\-me $d \Bigl( \ln^{-1}[\beta(t)] \Bigr)^3$. 
The second term of the left expression of equation (\ref{thermo}) may be identified with
 \begin{equation}
\Bigg|  \frac{p(t)}{c^2} d \Bigl( \ln^{-1}[\beta(t)]  \Bigr)^3 \Bigg| \propto  dW \, , 
 \end{equation}
where $dW$ denotes the elementary
 analytically  continued stress-energy, contained in the same volume $d \Bigl( \ln^{-1}[\beta(t)]\Bigr)^3.$ 
We then relate the terms of the left side of expression (\ref{thermo}) to the components of the fundamental thermodynamics relation for an infinitesimal reversible process, obtaining
\begin{eqnarray}
 dU   & =  & dQ + dW = T dS  - PdV \label{1lt}  \\
 & = &  \Bigg| d \Bigl[\rho(t) \Bigl( \ln^{-1}[\beta(t)] \Bigr)^3\Bigr]\Bigg|  + \Bigg| \frac{p(t)}{c^2} d \Bigl( \ln^{-1}[\beta(t)] \Bigr)^3 \Bigg| \, . \nonumber
\end{eqnarray}
In this expression, $dU$ represents the internal energy, $T$ is the absolute temperature, $S$ is the entropy, $P$ is the pressure, and $V$ is the volume of the analytically continued domain.
We conclude that the analytically continued energy-stress conservation law in the expanding branch cut universe (\ref{ECln}) obeys the first law of thermodynamics.


  \section{Complex conjugation of Friedmann's equations}

 The results presented in equations (\ref{eq1}) and (\ref{beta2}) indicate that complexifying the FWLR metric implies complexifying the time variable, $t$. 
In the following we proceed to the complex conjugation of the analytically continued FLRW metric~\citep{Zen2020}, with the non-zero components of the metric tensor expressed in terms of complex analytically continued and holomorphic variables, i.e.,  complex differentiable variables $r_{\xi}$ and $a_{\xi}(t)$ (see Table (\ref{Table1}). We adopt the nomenclature {\it cyclic time}  for the real part of ${\cal T}$ ($t = R e {\cal T}$) and 
{\it cosmological time}  for the imaginary part of ${\cal T}$ ($\tau = Im {\cal T}$)\footnote{We use the terminology {\it cyclic time} time associated with
oscillating cyclic-models of the universe as introduced by A. Friedmann in his seminal paper~\citep{Friedmann1922} and the terminology {\it cosmological time}  for the Minkowski imaginary time representation.}

\begin{table}
\caption{Components of the complex conjugated FLRW analytically continued metric.}
\begin{center}
  \begin{tabular}{@{}cc@{}}
 $ g^*_{00} = g^{*00}= 1$ & \\
 $g^*_{11}=-  \frac{a_\xi^{*2}(t^*)}{1-k\,r_\xi^{*2}}$  &  $g^{*11} =  -\left( \frac{a_\xi^{*2}(t^*)}{1-k\,r_\xi^{*2}}\,\right)^{\!-1}$ \\ 
   $g^*_{22}$ =   $-   r_\xi^{*2}\,a_\xi^{*2}(t^*)$ 
 & $g^{*22}  =  -\left( \, r_\xi^{*2}\,a_\xi^{*2}(t^*)\,\right)^{\!-1}$ \\ 
   $ g^*_{33} = -   r_\xi^{*2}\,a_\xi^{*2}(t^*)\,\sin^2\theta$ & $  g^{*33} =  -\left(  r_\xi^{*2}\,a_\xi^{*2}(t^*)\,\sin^2\theta \right)^{\!-1}$ 
  \label{Table1}
  \end{tabular}
  \end{center}
 \end{table}

The line element for the analytically continued metric stands as
\begin{eqnarray}
ds_\xi^{*2}\! =\! dt^{*2} \! -  a_\xi^{*2}(t^*)\left[\frac{dr_\xi^{*2}}{1-k\,r_\xi^{*2}}+r_\xi^{*2}(d\theta^2+\sin^2\theta\,d\phi^2\,)\right] . \nonumber \\
\end{eqnarray}
 The {\it analytically continued} 
Christoffel symbols which are different from zero are shown in Table (\ref{Table2}).
\begin{table}
\caption{Components of the complex conjugated FLRW analytically continued Christofell symbols.}
\begin{center}
  \begin{tabular}{@{}cc@{}}
$\Gamma^{*0}_{11} = \frac{a^*_\xi \dot{a}^*_{\xi}}{ 1- k r_\xi^{*2}}$ & 
$\Gamma^{*0}_{22} =  r_\xi^{*2} a^*_\xi \dot{a}^*_{\xi}$ \\
$\Gamma^{*0}_{33} =  r_\xi^{*2} a^*_\xi\, \dot{a}^*_{\xi}\,\sin^2\,\theta$ &
$\Gamma^{*1}_{11}  =   \frac{k\,r^*_\xi }{ 1- k\,r_\xi^{*2}}$ \\
$ \Gamma^{*1}_{22}  =    -r^*_\xi (1-k\, r_\xi^{*2})$ & 
$ \Gamma^{*1}_{33}  =    -r^*_\xi (1-k\, r_\xi^{*2})\,\sin^2\,\theta $ \\
$ \Gamma^{*2}_{33} = -\cos\theta\,\sin\theta$ & $\Gamma^{*2}_{12}  =  \Gamma^{*2}_{21}=\Gamma^{*3}_{13}=\Gamma^{*3}_{31}=\frac{1}{r_\xi} $ \\
$ \Gamma^{*3}_{23} = \Gamma^{*3}_{32}={\rm cotg}\,\theta$  &
$ \Gamma^{*1}_{01} =   \Gamma^{*1}_{10}= \Gamma^{*2}_{02}=\Gamma^{*2}_{20}$ \\
& $=\Gamma^{*3}_{03}=\Gamma^{*3}_{30}=\frac{\dot{a}^*_{\xi}}{a^*_\xi}$
  \label{Table2}
  \end{tabular}
  \end{center}
 \end{table}

The non-zero components of Einstein's mixed tensor $G^{*\mu}_{\phantom{x} \nu}$ are
\begin{eqnarray}
G^{*0}_{\phantom{x} 0}& = & -3  \left[\frac{\dot{a}_{\xi}^{*2}+k}{a_\xi^{*2}}\right]  ;  \nonumber \\
G^{*1}_{\phantom{x} 1} & = & G^{*2}_{\phantom{x} 2}  =  G^{*3}_{\phantom{x} 3} = 
- \left[\frac{2\ddot{a}^*_{\xi}}{a^*_\xi}+\frac{\dot{a}_{\xi}^{*2}+k }{a_\xi^{*2}}\right]. \label{gr}
\end{eqnarray}
The complex conjugated expression for the perfect fluid matter tensor of the universe stands as
\begin{eqnarray}
T_{\xi}^{*\mu\nu}=-p^*_{\xi}\,g_{\xi}^{*\mu\nu}+(p^*_{\xi}+\rho^*_{\xi})\,U_{\xi}^{*\mu}\,U_{\xi}^{*\nu}\,.
\end{eqnarray}

Combining these expressions, the analytically continued Friedmann's equations for the $\Lambda$CDM  ($\Lambda\neq 0$) model are
\begin{eqnarray}
H^*_{\xi}(t^*)&=&\frac{8\pi\,G}{3\,}\rho^*_{\xi}(t^*)-\frac{k\,}{a^{*2}_{\xi}(t^*)}+\frac{\Lambda_{\xi}\,}{3} \label{H*}
\\
2\frac{\ddot{a}^*_{\xi}}{a^*_{\xi}}&=&-{8\pi\,G}\,p^*_{\xi}(t^*)-H^*_{\xi}(t^*) -\frac{k\,}{a^{*2}_{\xi}}+\Lambda_{\xi}\,, \label{a*}
\end{eqnarray}
where $H^*_{\xi}(t^*)=\dot{a^{*2}_{\xi}}(t^*)/a^{*2}_{\xi}(t^*)$.

Following a similar previous technical procedure (see~\citet{Zen2020}), we 
arrive at the following complex conjugated Friedmann's-type  equations:
\begin{equation}
 \Biggl(\frac{\frac{d}{dt} \ln^{-1}(\beta^*(t^*))}{\ln^{-1}(\beta^*(t^*))} \Biggr)^2   =     \frac{8 \pi G}{3} \rho^*(t^*)
-  \frac{kc^2}{\ln^{-2}(\beta^*(t^*))} + \frac{1}{3} \Lambda^* ,   \label{NCFE1} 
\end{equation}
and
\begin{equation}
  \Biggl( \frac{\frac{d^2}{dt^{*2}} \ln^{-1}(\beta^*(t^*))}{\ln^{-1}(\beta^*(t^*)) } \Biggr)    =   - \frac{4 \pi G}{3} \Biggl(  \rho^*(t^*)  +  \frac{3}{c^2} p^*(t^*)  \Biggr)
+  \frac{1}{3} \Lambda^*  .   \label{NCFE2}
\end{equation}

The corresponding complex conjugated expression of the energy-stress conservation law in the expanding universe is given by
\begin{equation}
\frac{d \rho^*(t^*)}{d t^*} + 3 \Big(\rho^*(t^*) + \frac{p^*(t^*)}{c^2}  \Bigr) \Biggl( \frac{\frac{d}{d t^*} \ln^{-1}[\beta^*(t^*)]}{\ln^{-1}[\beta^*(t^*)]}\Biggr)  =  0 \, .  \label{CECln}
  \end{equation}%
Similar complex conjugated expressions for the previous cases for radiation-, matter-, and dark matter-dominated eras as well as for the conformal time can be obtained. 

\subsection{Tracing back the analytically continued FLRW metric}
 
Tracing back our results, 
the analytically continued FLRW metric stands out as
\begin{equation}
ds_{[\rm{ac}]}^2 \! = \! dt^2 \! - \! \ln^{-2}(\beta(t)) \Biggl[\!
\frac{dr^2}{\bigl(1 \! - \! kr^2(t) \bigr)}
\! + \! r^2(t) \Bigl(d \theta^2 \! + \! \sin^2 \theta d\phi^2 \Bigr) \! \label{FLRWac}   
\Biggr] ,
\end{equation}
with $r$ and $t$ representing space and time complex parameters and $k$ encoding the spatial curvature of the multi-composed universe, $k = -1, 0, 1$ for, respectively, negatively curved, flat or positively curved spatial hyper-surfaces analytically continued to the complex plane. 
The corresponding non-zero components of the metric tensor, expressed in terms of complex analytically continued and holomorphic variables, i.e.,  complex differentiable variables $r(t)$ and $\ln^{-1}[\beta(t)]$, are shown in Table (\ref{Table3})).

\begin{table}
\caption{Non-zero components of the metric tensor, expressed in terms of complex analytically continued and holomorphic variables, i.e.,  complex differentiable variables $r(t)$ and $\ln^{-1}[\beta(t)]$.}
\begin{center}
  \begin{tabular}{@{}cc@{}}
$g_{00}= g^{00}=1$ &\\
 $g_{11} = -  \frac{\ln^{-2}[\beta(t)]}{1-k\,r^{2}(t)}$ &
 $g^{11}  =  -\left(\! \frac{\ln^{-2}[\beta(t)]}{1-k r^{2}(t)} \! \right)^{\!\!-1}$ \\
  $g_{22}  =  -  r^{2}(t)  \ln^{-2}[\beta(t)]$  & 
$g^{22}  =  -\left(r^{2}(t) \, \ln^{-2}[\beta(t)]\right)^{\!\!-1} $ \\
 $g_{33} = -   r^{2}(t)  \ln^{-2}[\beta(t)]\sin^2\theta$  & 
$g^{33}  =  -\left(\!  r^{2}(t) \ln^{-2}[\beta(t)]  \sin^2\theta \! \right)^{\!\!-1}\!\!\!.$   
  \label{Table3}
  \end{tabular}
  \end{center}
 \end{table}

The {\it analytically continued} 
Christoffel symbols which are different from zero are:
\begin{eqnarray}
\Gamma^{0}_{11}& =& \frac{\ln^{-1}[\beta(t)] \, \frac{d}{dt}\ln^{-1}[\beta(t)]}{ 1- k\,r^{2}(t)}
; \nonumber \\
\Gamma^{0}_{22} &= & r^{2}(t) \, \ln^{-1}[\beta(t)] \, \frac{d}{dt} \ln^{-1}[\beta(t)]
; \nonumber \\
\Gamma^{0}_{33} & = & r^{2}(t) \, \ln^{-1}[\beta(t)] \, \Bigl( \frac{d}{dt}\ln^{-1}[\beta(t)]\Bigr) \, \sin^2\,\theta
 ;   \nonumber \\
\Gamma^{1}_{11} & = &  \frac{k\,r(t)}{ 1- k\,r^{2}(t)}
 ; \,\,\,
\Gamma^{1}_{22}   =    -r(t) (1-k\, r^{2}(t))
 ;  \nonumber \\
\Gamma^{1}_{33} & = &   -r(t) (1-k\, r^{2}(t))\,\sin^2\,\theta; \,\,\,
\Gamma^{2}_{33}=-\cos\theta\,\sin\theta ; 
\nonumber \\
\Gamma^{1}_{01}& =  & \Gamma^{1}_{10}= \Gamma^{2}_{02}=\Gamma^{2}_{20}=\Gamma^{3}_{03}=\Gamma^{3}_{30}=\frac{\frac{d}{dt}\ln^{-1}[\beta(t)]}{\ln^{-1}[\beta(t)]} \, ; 
\nonumber
\\
\Gamma^{2}_{12} & = & \Gamma^{2}_{21}=\Gamma^{3}_{13}=\Gamma^{3}_{31}=\frac{1}{r(t)}
\, ; \,\,\,
\Gamma^{3}_{23}=\Gamma^{3}_{32}={\rm cotg}\,\theta\,. \label{Gammaac} \nonumber \\
\end{eqnarray}
The non-zero components of Einstein's mixed tensor $G^{*\mu}_{\phantom{x} \nu}$ are
\begin{equation}
 G^{0}_{\phantom{x} 0} =  -3  \left[\frac{\Bigr(\frac{d}{dt} \ln^{-1}[\beta(t)]\Bigl)^{2}+k}{\Bigr( \ln^{-1}[\beta(t)]\Bigl)^{2}}\right]  ;  \label{G0}  
\end{equation}
\begin{equation}
G^{1}_{\phantom{x} 1}  \!= \! G^{2}_{\phantom{x} 2}  \! =\!  G^{3}_{\phantom{x} 3} \!=\! 
- \left[\frac{2\frac{d^2}{dt^2} \ln^{-1}[\beta(t)]}{\ln^{-1}[\beta(t)]}+\frac{\Bigr(\frac{d}{dt} \ln^{-1}[\beta(t)]\Bigl)^{2}+k }{\ln^{-2}(\beta(t))}\right].  \label{G123}
\end{equation}

\section{Results and Conclusions}

In synthesis, the scale factor $\ln^{-1}[\beta(t)]$ is a  dimensionless scalar complex time-dependent function and represents the relative expansion of the universe, relating the co-moving distances for an expanding universe with the distances in an arbitrary spacetime referential frame. And its inverse, $\ln[\beta(t)]$, represents the corresponding scaling factor in time. The limitations imposed by the presence of singularities in general relativity are replaced in this type of treatment by functions that behave continuously in the real domain but are complemented by discontinuity jumps on the imaginary axis. This type of treatment may represent a technical alternative for overcoming the undesirable presence of singularities in general relativity in the regime of strong gravity and/or high spacetime curvatures. Additionally, as we will see later, in a forthcoming contribution~\citep{ZenB}, the (apparent) formal inconvenience of complexifying the FWLR metric has a relatively simple solution if we associate its temporal dependence with a Wick rotation to Euclidean space, allowing this way
 the realisation of the (new) scale factor of general relativity. 
 
The present formalism presents similarity with quantum bouncing models, regarding a period of expansion followed by a bounce and a smooth transition to a period of contraction in the evolutionary universe. Although diverse, bouncing models assume in general a mechanism (or trigger)
 to keep the bouncing phase stable which could be associated for example with quantum fluctuations. In this kind of quantum bouncing models, the contraction phase amplifies quantum fluctuations and could serve as a trigger for the expansion phase (see for instance \citet{Novello2008}).
 
 Our present formulation in turn allows  a thermodynamically consistent kind of `classical' {\it tunnelling} between the remote past of the universe's evolutionary process in which the spacetime fabric develops continuously along Riemann sheets that circumvent a branch cut, thus avoiding discontinuities in the general relativity equations. In the {\it branch cut universe}, the regularisation functions $\chi(t)$ mimic the underlying mechanism of keeping stable the classic transition from the negative complex cosmological time sector $t_C$, prior to any conception of primordial singularity, to the positive cosmological time sector.  
  In a following contribution~\citep{ZenB}, we sketched a quantum formulation which represents a kind of quantum  tunnelling between the contraction and the expansion phases, an effect quite similar to the corresponding tunnelling effect in which a particle may penetrate through a potential energy barrier that is higher in energy than the particle's kinetic energy, that is, an effect of a purely quantum nature. 
  
\section{Acknowledgements}
P:O:H: acknowledges financial support from PAPIIT-DGAPA (IN100421). The authors would like to thank the referees for valuable comments.

\appendix

\section{Ages in a branch cut universe}\label{a}

\subsection{Radiation-dominated era: perfect fluid approximation}

From the first Friedmann's equation, extended to the complex plane (\ref{NFE1}), in case $\Lambda = k = 0$, for the radiation-dominated era, with $p = \frac{1}{3} \rho(t)$ we get
\begin{equation}
\frac{\Bigl(\frac{d}{dt} \ln^{-1}[\beta(t)] \Bigr)}{\Bigl( \ln^{-1}[\beta(t)] \Bigr)}  \! = \! \sqrt{ \frac{8 \pi G \rho(t)}{3} }  
 \! = \! \frac{1}{\ln^{-2}[\beta(t)] \ln^{\, 2}(\beta_0)}  \sqrt{ \frac{8 \pi G \rho_0}{3}}  \, , \label{FFE}
\end{equation}
resulting in
\begin{equation}
\to \, \, \, \ln^{-1}[\beta(t)]  = \sqrt{\ln^{-2}[\beta(t_P)]   
+ \frac{1}{ \ln^{2}(\beta_0)} \sqrt{\frac{2 \pi G \rho_0}{3}} \Bigl( t - t_P  \Bigr) 
 } \label{eq1} \, . 
\end{equation}
From this equation, we can isolate the $\beta(t)$ parameter: 
\begin{equation}
\beta(t)   =    \ln\sqrt{  \ln^{-2}[\beta(t_P)]   
+ \frac{1}{ \ln^{2}(\beta_0)} \sqrt{\frac{2 \pi G \rho_0}{3}} \Bigl( t - t_P \Bigr) }. \label{beta}
\end{equation}
A generalisation of this result for $k \neq 0$  then holds:
\begin{equation}
\Biggl( \frac{\frac{d}{dt} \ln^{-1}[\beta(t)]}{\ln^{-1}[\beta(t)]  } \Biggr)^2       =    \frac{8 \pi G}{3} \frac{\rho_0}{\ln^{4}(\beta_0) \ln^{-4}[\beta(t)]}
 -  \frac{kc^2}{\ln^{-2}[\beta(t)]}  \, . 
\end{equation}
From this expression, the analytically continued conformal time 
 is given as:
\begin{eqnarray}
\eta(t)-\eta(t_P) & =& 
   \int_{t_P}^{t} \frac{dt}{\ln^{-1}[\beta(t)]}   \label{acct} \\
 & = &  \int_{\ln^{-1}(\beta(t_P))}^{\ln^{-1}[\beta(t)]} \frac{d \ln^{-1}[\beta(t)]}{\sqrt{ \frac{8 \pi G}{3} \rho_0 \ln^{-4}(\beta_0)
-  kc^2 \ln^{-2}[\beta(t)]}} \, . \nonumber
\end{eqnarray}
From this equation,
for $k = 1$,  we obtain:
\begin{equation}
 \eta(t) - \eta(t_P) = \frac{1}{c} \sin^{-1}  \Biggl(\!  \frac{ \ln^{-1}[\beta(t)]}{\sqrt{ \frac{8 \pi G}{3 c^2} \rho_0 \ln^{-4}(\beta_0) }} \! \Biggr) \Bigg|_{\ln^{-1}[\beta(t_P)]}^{\ln^{-1}[\beta(t)]} \, . 
\end{equation}
For $ k = -1$, $\eta(t) - \eta(t_P)$ becomes:
\begin{equation}
 \eta(t) - \eta(t_P) = \frac{1}{c} \sinh^{-1}  \Biggl(\!  \frac{ \ln^{-1}[\beta(t)]}{\sqrt{ \frac{8 \pi G}{3 c^2} \rho_0 \ln^{-4}(\beta_0) }} \! \Biggr) \Bigg|_{\ln^{-1}[\beta(t_P)]}^{\ln^{-1}[\beta(t)]} \, . 
\end{equation}


\subsection{Matter-dominated era: dust approximation}

In the matter-dominated era of the universe, from the first Friedmann's equation (\ref{NFE1}), extended to the complex plane, in case $\Lambda = k = 0$, with $p = 0$ 
we get

\begin{equation}
\frac{\Bigl(\frac{d}{dt} \ln^{-1}[\beta(t)] \Bigr)}{\Bigl( \ln^{-1}[\beta(t)] \Bigr)}   =  \frac{1}{\ln^{-3/2}[\beta(t)] \ln^{3/2}(\beta_0)}  \sqrt{ \frac{8 \pi G \rho_0}{3}}  \, . \label{FFE}
\end{equation}
From this equation we obtain
\begin{equation} 
 \int_{\ln^{-1}(\beta(t_P))}^{\ln^{-1}[\beta(t)]} \!\!\!\!\!\!\!
\Bigl(\! \ln^{-1}[\beta(t)] \! \Bigr)^{\!1/2} \! \Bigl( \! \frac{d}{dt} \ln^{-1}[\beta(t)] \! \Bigr) dt  =    \sqrt{ \frac{8 \pi G \rho_0}{3 \ln^{3}(\beta_0)}} \!\! \int_{t_P}^{t}  dt .
 \end{equation}
Therefore,
 \begin{equation}
\ln^{-1}[\beta(t)]  \! = \!\! \sqrt[2/3]{ \ln^{-3/2}[\beta(t_P)]   
 \!+ \! \frac{1}{ \ln^{\, 3/2}(\beta_0)}  \sqrt{6 \pi G \rho_0} \Bigl(\! t - t_P \! \Bigr)
 } .  \label{eq2}
\end{equation}
From this expression we can isolate the $\beta(t)$ parameter
\begin{equation}
\beta(t)    \! =  \!  \ln^{-1}\Biggl[ \! \sqrt[2/3]{ \ln^{-3/2}(\beta(t_P))   
 +  \frac{1}{ \ln^{\, 3/2}(\beta_0)}  \sqrt{6 \pi G \rho_0} \Bigl(t - t_P \Bigr)
 }  \Biggr].  \label{beta2}
\end{equation}

A generalisation of this result for $k \neq 0$  then holds:
\begin{equation}
\Biggl( \frac{\frac{d}{dt} \ln^{-1}[\beta(t)]}{\ln^{-1}[\beta(t)]  } \Biggr)^2    
    =     \frac{8 \pi G}{3}\frac{\rho_0}{\ln^{3}(\beta_0) \ln^{-3}[\beta(t)]}
 -  \frac{kc^2}{\ln^{-2}[\beta(t)]}  \, . 
\end{equation}
For $k = 1$, from this expression, $
\eta(t) - \eta(t_P)$ is given as:
\begin{equation} 
\eta(t) - \eta(t_P) =  \frac{1}{c} \sin^{-1} \Biggl(\!  \frac{ \ln^{-1}[\beta(t)] -  \frac{4 \pi G}{3c^2} \rho_0 \ln^{-3}(\beta_0)}{\frac{4 \pi G}{3c^2} \rho_0 \ln^{-3}(\beta_0)}   \!        \Biggr) \! \Bigg|_{\ln^{-1}[\beta(t_P)]}^{\ln^{-1}[\beta(t)]} .
\end{equation}
Similarly, for $ k = -1$, we obtain:
\begin{equation}
\eta(t) - \eta(t_P) =  \frac{1}{c} cosh^{-1}  \Biggl(  \frac{ \ln^{-1}[\beta(t)] + \frac{4 \pi G}{3c^2} \rho_0 \ln^{-3}(\beta_0)}{\frac{4 \pi G}{3c^2} \rho_0 \ln^{-3}(\beta_0)}           \Biggr) \Bigg|_{\ln^{-1}[\beta(t_P)]}^{\ln^{-1}[\beta(t)]} .  
\end{equation}

\bibliography{ZenA}%

\end{document}